\documentclass[12pt]{article}
\usepackage{scicite}
\usepackage{graphicx}
\usepackage{amssymb}
\usepackage{amsthm,lineno}
\usepackage{float}
\usepackage{amsmath}
\usepackage{booktabs}
\usepackage{braket}
\usepackage{xcolor}
\usepackage{subfigure}

\usepackage{times}
\usepackage{array}
\usepackage{xcolor}
\def\noteC#1{\textbf{\color{green}}} %part that had been cancelled
\usepackage{amssymb, amsmath}

\usepackage{setspace}

\newenvironment{sciabstract}{%
\begin{quote} \bf}
{\end{quote}}

\newcounter{lastnote}

\def\noteC#1{\textbf{\color{green}}} %part that had been cancelled
\title{Quantum simulation of molecular spectroscopy in  trapped-ion device}

\author
{Yangchao Shen$^{1}$, Joonsuk Huh$^{2*}$, Yao Lu$^{1}$, Junhua Zhang$^{1}$, \\
Kuan Zhang$^{1}$, Shuaining Zhang$^{1}$ and Kihwan Kim$^{1\dagger}$\\
\\
\normalsize{$^{1}$Center for Quantum
Information, IIIS, Tsinghua University, Beijing, P. R. China,}\\
\normalsize{$^{2}$Department of Chemistry, Sungkyunkwan University, Suwon 440-746, Korea}\\
\\
\normalsize{To whom correspondence should be addressed;} 
\\
\normalsize{E-mails:  $^\ast$joonsukhuh@gmail.com and $^\dagger$kimkihwan@mail.tsinghua.edu.cn}
}

\date{}

\begin{document} 
%\singlespacing

%\baselineskip24pt
%\baselineskip36pt

\maketitle 

%\begin{multicols}{3}
\begin{sciabstract}
%Boson sampling is designed to simulate a classically not accessible probability distribution of multi-photons scattered through a photonic network. The classical simulation of the multi-photon distribution is intractable as the problem size increases. 

Molecules are the most demanding quantum systems to be simulated by quantum computers because of their complexity and the emergent role of quantum nature. The recent theoretical proposal of Huh et al. (Nature Photon., 9, 615 (2015)) showed that a multi-photon network with a Gaussian input state can simulate a molecular spectroscopic process. Here, we report the first experimental demonstration of molecular vibrational spectroscopy of SO$_{2}$ with a trapped-ion system. In our realization, the molecular scattering operation is decomposed to a series of elementary quantum optical operations, which are implemented through Raman laser beams, resulting in a multimode Gaussian (Bogoliubov) transformation. The molecular spectroscopic signal is reconstructed from the collective projection measurements on phonon modes of the trapped-ion system. Our experimental demonstration would pave the way to large-scale molecular quantum simulations, which are classically intractable. 
\\\\One Sentence Summary: Gaussian boson sampling for photoelectron spectra of sulfur dioxide has been performed using a trapped-ion quantum simulator.
\end{sciabstract}
%]

% In setting up this template for *Science* papers, we've used both
% the \section* command and the \paragraph* command for topical
% divisions.  Which you use will of course depend on the type of paper
% you're writing.  Review Articles tend to have displayed headings, for
% which \section* is more appropriate; Research Articles, when they have
% formal topical divisions at all, tend to signal them with bold text
% that runs into the paragraph, for which \paragraph* is the right
% choice.  Either way, use the asterisk (*) modifier, as shown, to

%\section*{Introduction}
%\section*{Results}
%Reports (up to ~2500 words including references, notes and captions or ~3 printed pages) present important new research results of broad significance. Reports should include an abstract, an introductory paragraph, up to four figures or tables, and about 30 references. Materials and Methods should usually be included in supplementary materials, which should also include information needed to support the paper's conclusions.

%Main Text: In general, this should include a brief (1-2 paragraph) introduction, followed by a statement of the specific scope of the study, followed by results and then interpretations.  Please avoid statements of future work or claims of priority, and avoid repeating the conclusions at the end.  

%Boson sampling problem is a strong candidate, which could uncover the quantum supremacy~\cite{Aaronson2011}. 
Boson sampling was originally proposed to simulate a classically intractable multiphoton distribution of indistinguishable photons scattered by beam splitters and phase shifters~\cite{Aaronson2011}. The successful small-scale experimental implementations
~\cite{Spring2013,Broome2013,Crespi2013,Tillmann2013} of the boson sampling paved the way for quantum simulators regardless of the lack in its obvious practical applications.
Alternatively, a modified boson sampling with Gaussian input states, such as thermal and squeezed vacuum states, has been discussed in the computational complexity perspective, classifying that the boson sampling with squeezed states as a classically hard problem~\cite{Lund2014,rahimi2015}. 
Recently, Huh et al.~\cite{huh2015,huh2016} proposed  Gaussian boson sampling as a practical application of the boson sampling in connecting the output to the molecular vibronic (vibrational+electronic) spectroscopy. Essentially, boson sampling turned out to be able to not only answer the computational complexity questions but also have a practical application. 

\begin{figure}[htbp]
\centering
\includegraphics[width=\textwidth]{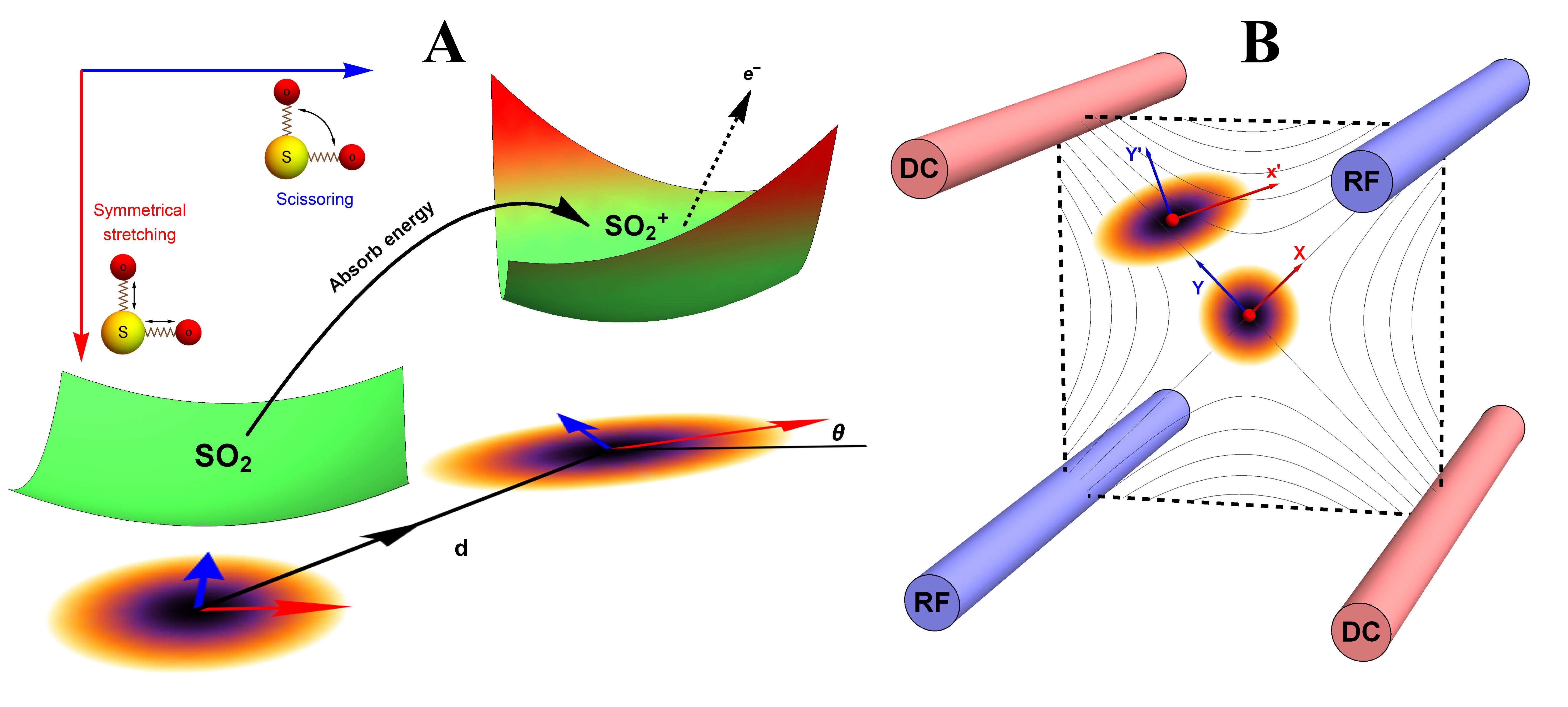}
\caption{\textbf{Pictorial description of the photoelectron spectrocopy of SO$_2$ and the trapped-ion simulator.} (A) The photoelectron process of SO$_2$ $\rightarrow$ SO$_2^+$. The molecule is initially at the vibrational ground state of the symmetrical stretching and scissoring modes. After absorbing a photon, an electron is removed from the molecule and the molecule finds a new equilibrium structure for SO$_2^+$, where the new potential energy surface is displaced, squeezed, and rotated from the original one. The transition of SO$_2^-$ $\rightarrow$ SO$_2$ can be described by a similar manner. (B) The trapped-ion simulator performing the Gaussian transformation for the molecular vibronic spectroscopy. The two vibrational modes of SO$_2$ are mapped to two radial modes (X and Y) of a single trapped-ion. %\note{Joon: what is radial mode?} 
The photoelectron process is simulated by applying series of quantum optical operations, which are implemented by Raman laser beams (see SM.C.). Generally, the photoelectron process of more complicated molecules with $N$ vibrational modes can be mapped to the collective motional modes of $N$ ions with the similar operations by Raman laser beams. %\note{Joon: Confused, N or N/2? give a reference if available.} 
}
\label{Setup}
\end{figure}

The molecular vibronic spectroscopy carries the vibrational transitions between nuclear manifolds belonging to two electronic states of a molecule~\cite{jankowiak:2007,huh2015}, as shown in Fig.~\ref{Setup}(A). Upon the electronic transition, a molecule undergoes structural deformation, vibrational frequency changes and rotation of normal modes; within a harmonic approximation to the electronic potential energy surfaces, these are equivalent to the displacement ($\hat{D}$), squeezing ($\hat{S}$) and rotation ($\hat{R}$) operations in quantum optics, respectively. 
The (mass-weighted) normal coordinates of initial ($\mathbf{Q}$) and final ($\mathbf{Q}'$) states are related linearly as $\mathbf{Q}'=\mathbf{U}\mathbf{Q}+\mathbf{d}$, where $\mathbf{U}$ is called the Duschinsky rotation matrix and $\mathbf{d}$ is a displacement vector of the multidimensional harmonic oscillators in the mass-weighted coordinate, the corresponding dimensionless displacement vector $\boldsymbol{\delta}$ for the quantum optical operation is used later~\cite{jankowiak:2007}.   
As a result, the molecule performs a multi-mode Bogoliubov transformation~\cite{Braunstein2005} between the 
(vibrational) boson operators of the initial and final electronic states~\cite{huh2015,huh2016}. The probability distribution regarding a given molecular vibronic transition frequency ($\omega_{\mathrm{v}}$) at zero Kelvin, that is, spectrum (Franck-Condon profile), is read as a Fermi's golden rule for a unitary Gaussian operator $\hat{U}_{\mathrm{Dok}}$~\cite{doktorov:1977,jankowiak:2007,huh2015}, 
\begin{align}
F(\omega_{\mathrm{v}})=\sum_{\mathbf{m}=\mathbf{0}}^{\boldsymbol{\infty}}\vert\langle\mathbf{m}\vert\hat{U}_{\mathrm{Dok}}\vert\mathbf{0}\rangle\vert^{2}\delta(\Delta_{\omega_{\mathrm{v}}})
\label{eq:FCW}
\end{align}
where $\Delta_{\omega_{\mathrm{v}}}=\omega_{0-0}+\omega_{\mathrm{v}}-(\sum_{k=1}^{M}m_{k}\omega_{k}')$, with the $k$-th vibrational frequency ($\omega_{k}'$) of a molecule in the final electronic state ($\omega_{k}$ belongs to the initial electronic state). The constant off-set frequency $\omega_{0-0}$, which includes the electronic transition and the zero-point vibrational transition, is set to be zero in this report without losing the generality.      
%and $F(\omega_{\mathrm{v}})$ is usually called Franck-Condon factor weighted density of states,
$\vert\mathbf{0}\rangle=\vert0_{1},\ldots,0_{M}\rangle$ and $\vert\mathbf{m}\rangle=\vert m_{1},\ldots,m_{M}\rangle$ are the initial and final $M$-dimensional Fock states, respectively. 
 
%In Eq.~\ref{eq:FCW}, the transition probabilities (Franck-Condon factors), are weighted by a density of states at the vibrational transition frequency $\omega_{\mathrm{v}}=\sum_{k=1}^{M}m_{k}\omega_{k}'$. 
% The Gaussian operator $\hat{U}_{\mathrm{Dok}}$ performs a multi-mode Bogoliubov transformation to the initial bosonic operators $\{\hat{a}_{k}\}$ and $\{\hat{a}_{k}^{\dagger}\}$, where the annihilation and creation operators satisfy $[\hat{a}_{k},\hat{a}_{l}^{\dagger}]=\delta_{k,l}$. 
% That is, $\hat{U}_{\mathrm{Dok}}^{\dagger}\hat{a}_{k}^{\dagger}\hat{U}_{\mathrm{Dok}}=\sum_{l=1}^{M}X_{k,l}\hat{a}_{l}+\sum_{l=1}^{M}Y_{k,l}\hat{a}_{l}^{\dagger}+z_{k}$, where the $M$-dimensional square matrices satisfy $XX^{\dagger}-YY^{\dagger}=I$ and $XY^{\mathrm{t}}=YX^{\mathrm{t}}$~\cite{Braunstein2005} and $z_{k}$ is a linear shift, which is related to the displacement operator. 
%The Gaussian unitary operator $\hat{U}_{\mathrm{Dok}}$ can be decomposed into sequential quantum optical operations~\cite{doktorov:1977,Braunstein2005,huh2015};  Huh et al.~\cite{huh2015} proposed a Gaussian boson sampling experiment: The prepared squeezed coherent states pass through a photon network and the output photon numbers are measured to reconstruct the Franck-Condon profile (FCP) in Eq.~\ref{eq:FCW}. 
Doktorov et al.~\cite{doktorov:1977} decomposed $\hat{U}_{\mathrm{Dok}}$ in terms of the elementary quantum optical operators as follows:   
\begin{align}
\hat{U}_{\mathrm{Dok}}=\hat{D}_N(\boldsymbol{\delta})\hat{S}^{\dagger}_N(\boldsymbol{\zeta'}) \hat{R}_N(\mathbf{U})\hat{S}_N(\boldsymbol{\zeta})
\label{eq:Doktorov3}
\end{align}
where $\hat{D}_N,\hat{S}_N$ and $\hat{R}_N$ are the $N$-mode operators of displacement, squeezing and rotation \cite{Ma1990} (see also Supplementary Materials (SM), SM.A.); $\boldsymbol{\delta}(=\boldsymbol{\zeta}'\mathbf{d}/\sqrt{2\hbar})$ is a (dimensionless) molecular displacement vector, $\boldsymbol{\zeta}=\mathrm{diag}(\ln \sqrt{\omega_{1}},$ $\ldots,\ln \sqrt{\omega_{N}} )$ and $\boldsymbol{\zeta'}=\mathrm{diag}(\ln \sqrt{\omega_{1}'},\ldots,\ln \sqrt{\omega_{N}'})$ are diagnoal matrices of the squeezing parameters, and $\boldsymbol{U}$ is a unitary rotation matrix. 
%$\{\omega_{k}\}$ is the $k$-th harmonic frequencies of initial electronic states of the molecule.
The actions of the quantum optical operators are defined in Ref.~\cite{huh2015}. Therefore, the sequential operations of the quantum optical operators in Eq.~\eqref{eq:Doktorov3} to the vacuum state and the measurement in Fock basis, as  in Eq.~\eqref{eq:FCW}~\cite{huh2015}, can simulate the Franck-Condon profile.   
% \begin{align}
% &\hat{D}_{\boldsymbol{\delta}/\sqrt{2}}^{\dagger}\mathbf{\hat{a}}^{\dagger}\hat{D}_{\boldsymbol{\delta}/\sqrt{2}}=\mathbf{\hat{a}}^{\dagger}+\tfrac{1}{\sqrt{2}}\boldsymbol{\delta} \nonumber \\
% &\hat{S}_{\Omega'}\mathbf{\hat{a}}^{\dagger}\hat{S}_{\Omega'}^{\dagger}=\tfrac{1}{2}\left(\Omega'+\Omega'^{-1}\right)\mathbf{\hat{a}}^{\dagger}
% +\tfrac{1}{2}\left(\Omega'-\Omega'^{-1}\right)\mathbf{\hat{a}} \nonumber \\
% &\hat{S}_{\Omega}^{\dagger}\mathbf{\hat{a}}^{\dagger}\hat{S}_{\Omega}=\tfrac{1}{2}\left(\Omega+\Omega^{-1}\right)\mathbf{\hat{a}}^{\dagger}
% -\tfrac{1}{2}\left(\Omega-\Omega^{-1}\right)\mathbf{\hat{a}}
% \end{align}
% for the boson creation and annihilation operator vectors $\mathbf{\hat{a}}^{\dagger}$ and $\mathbf{\hat{a}}$
% \begin{align}
% &\mathbf{J}=\boldsymbol{\Omega}'\mathbf{U}\boldsymbol{\Omega}^{-1}, \quad
% \boldsymbol{\delta}=\hbar^{-\tfrac{1}{2}}\boldsymbol{\Omega}'\mathbf{d}, \nonumber \\
% &\boldsymbol{\Omega}'=\mathrm{diag}(\sqrt{\omega_{1}'},\ldots,\sqrt{\omega_{N}'}), \nonumber \\
% &\boldsymbol{\Omega}=\mathrm{diag}(\sqrt{\omega_{1}},\ldots,\sqrt{\omega_{N}}) \, .
% \label{eq:parameters}
% \end{align}

Not only the original boson sampling but also the Gaussian boson sampling for the vibronic spectrum, however, are challenging in an optical system \cite{Spring2013,Broome2013,Crespi2013,Tillmann2013} because of the difficulties in preparing the initial states: single Fock states for the original boson sampling and squeezed coherent states for the molecular simulation. Non-optical boson sampling devices, such as trapped-ion~\cite{Lau2012A,Shen2014} and superconducting circuit ~\cite{Peropadre2015}, have been suggested theoretically for the scalable boson sampling machine to overcome the difficulties of the optical implementation in preparing the single photon states. Moreover, these non-optical devices can handle the squeezed states with relative ease.  
%Moreover, these non-optical devices can handle relatively easily the required Gaussian input states (squeezed coherent states) for the molecular spectroscopy. Especially, the necessary experimental techniques of the molecular simulation are already available for the trapped-ion device; the state preparation \cite{Monroe1995,Roos1999}, quantum operations \cite{Toyoda2015,Meekhof1996} and state detection \cite{Watanabe2011,An2014}.
In this report, we present the first quantum simulation of molecular vibronic spectroscopy with a particular example of photoelectron spectroscopy of sulfur dioxide (SO$_{2}$)~\cite{nimlos1986,Lee2009}. 

Fig.~\ref{Setup}(B) schematically illustrates quantum optical operations of Eq.~\eqref{eq:Doktorov3} in the trapped-ion device for the molecular vibronic spectroscopy of SO$_{2}$. Our trapped-ion simulation is performed using a single $^{171}\mathrm{Yb}^+$ ion confined in the 3-dimensional harmonic potential generated by the four-rod trap. The two radial phonon modes (X and Y) of an ion, with the trap frequencies $\omega_\mathrm{X}=(2\pi)2.4~\mathrm{MHz}$ and $\omega_\mathrm{Y}=(2\pi)1.9~\mathrm{MHz}$, map the two vibrational modes of the molecule. After the mapping of the Hilbert space between the real molecule and simulator is established, the molecular spectroscopy is simulated through the following procedure: (i) the ion is first initialized in the motional  ground state, (ii) the quantum optical operations in Eq.~\eqref{eq:Doktorov3} are then sequentially applied, and (iii) finally, the vibronic spectrum is constructed using the collective projection measurements on the transformed state. 

Accordingly, for the first step of the molecular spectroscopy simulation, we prepare the ion in the ground state $\ket{n_{\rm X}=0,n_{\rm Y}=0}$ by the Doppler cooling and the resolved sideband cooling methods~\cite{Monroe1995,Roos1999}. Next, we perform the required displacement, squeezing and rotation operations by converting the molecular parameters to the corresponding device parameters. The molecular parameters $\boldsymbol{\delta}, \boldsymbol{\zeta'}, \boldsymbol{U}$ and $\boldsymbol{\zeta}$ can be obtained via conventional quantum chemical calculations with available program packages (e.g., Ref.~\cite{g16}). See SM.B., for the details of the parameter conversion for SO$_2$. 
%The conversion process also bridges the difference of the actual absolute frequencies between molecule and simultor.

\begin{figure}[htbp]
\centering
\includegraphics[width=\textwidth]{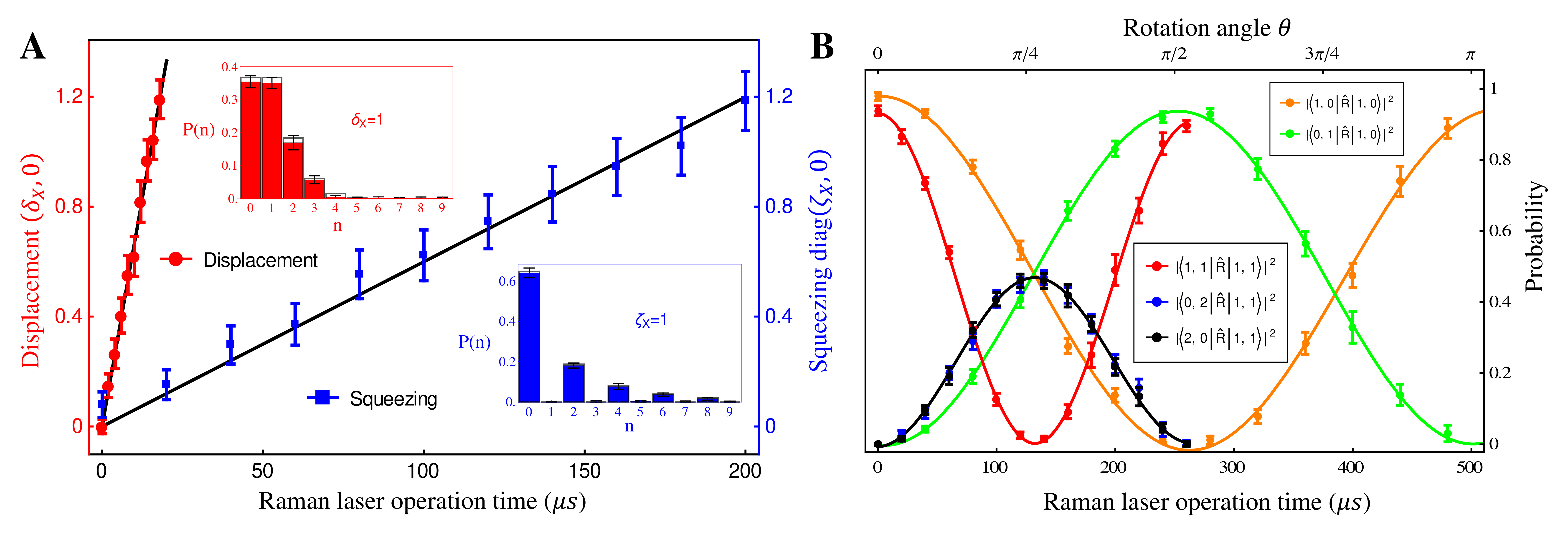}
\caption{\textbf{Trapped-ion demonstration of quantum optical operations; $\hat{D}, \hat{S},$ and $\hat{R}$.} (A) Displacement $\boldsymbol{\delta}=(\delta_{\rm X},0)$ (red) and squeezing $\boldsymbol{\zeta}={\rm diag}(\zeta_{\rm X},0)$ (blue) of mode X are controlled by the duration of Raman laser beams. The insets show the measured phonon distribution for $\delta_{\rm X}=1$ and for the squeezing parameter of $\zeta_{\rm X}=1$. (B) The evolution of rotation operation between mode X and Y. Here all the operations are implemented by Raman laser beams. The dots represent the experimental data and the lines are obtained by fitting. The error bars stand for 95$\%$ confidence level.}
\label{Operations}
\end{figure}

The quantum optical operations (displacement $\hat{D}$, squeezing $\hat{S}$ and rotation $\hat{R}$) are implemented by the $\sigma_+$-polarized Raman laser beams from a pico-second pulse laser with a wavelength of 375 nm (see SM.C.). 
In the trapped-ion experiment, the quantum optical operations with the desired parameters can be performed by controlling the duration and relative phase of the Raman beams. Therefore, the different Raman laser beams with frequencies of $\mathrm{\omega_X}(\mathrm{\omega_Y})$, $2\mathrm{\omega_X}(2\mathrm{\omega_Y})$ and $\mathrm{\omega_X-\omega_Y}$ are assigned for displacement, squeezing and rotation operations, respectively \cite{Meekhof1996,Toyoda2015}.  Fig.~\ref{Operations}(A) shows the performance of the experimental displacement $\hat{D}_2(\boldsymbol{\delta})=\hat{D}(\delta_{\rm X},0)=e^{\delta_{\rm X} \hat{a}_{\rm X}^\dag-\delta_{\rm X}^{*}\hat{a}_{\rm X}}$ and squeezing $\hat{S}_2(\boldsymbol{\zeta})=\hat{S}({\rm diag}(\zeta_{\rm X},0))=e^{\frac{1}{2}(\zeta_{\rm X}^{*}\hat{a}_{\rm X}\hat{a}_{\rm X}-\zeta_{\rm X}\hat{a}_{\rm X}^\dag\hat{a}_{\rm X}^\dag)}$ operations, where $a_{\rm X}$ and $a^\dag_{\rm X}$ are the annihilation and creation operators of bosonic mode X, respectively. The amount of the displacement $\alpha$ and the squeezing parameter $\zeta$ are controlled by the duration of the corresponding Raman beams with the rates of 0.066 $\mu s^{-1}$ and 0.006 $\mu s^{-1}$, respectively. We examine the trapped-ion implementation of the rotation operation $\hat{R}_2(\boldsymbol{U})=\hat{R}(\theta)=e^{\theta(\hat{a}_{\rm X}^\dag \hat{a}_{\rm Y}-\hat{a}_{\rm X}\hat{a}_{\rm Y}^\dag)}$ between modes X and Y with two sets of initial states, as indicated in Fig.~\ref{Operations}(B). The rotation angle $\theta$ is also controlled by the duration of the operation with a rate of 0.006 rad  $\mu s^{-1}$. The oscillations in Fig.~\ref{Operations}(B) of the initial state $\mathrm{\ket{n_X=1,n_Y=0}}$ (orange and green) are twice slower than those from state $\mathrm{\ket{1,1}}$ (blue, black and red), as expected. We note that at $t=131 \mu s$, the near zero probability of $\bra{1,1} \hat{R} \ket{1,1}$ originates from the Hong-Ou-Mandel interference \cite{Toyoda2015}.

\begin{figure}[htbp]
\centering
\includegraphics[width=\textwidth]{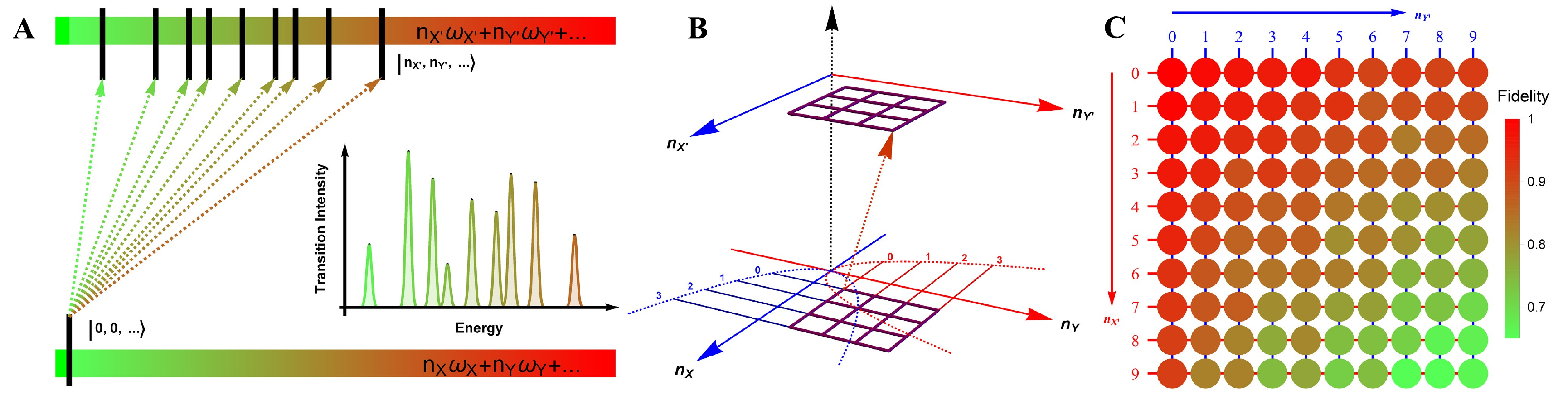}
\caption{\textbf{Construction scheme for the Franck-Condon profile of the photoelectron process with the trapped-ion simulator.}  (A) Generic diagram for molecular transition process at $T=0$ K. The lower bar indicates the initial state and the upper bar shows the final states after the process. The vibronic spectrum is constructed by measuring the transition probabilities from $\mathrm{\ket{n_X=0,n_y=0,...}}$ to  $\mathrm{\ket{n_{X'},n_{Y'},...}}$.  (B) The transition process of a molecule in the two-dimensional Fock space. The process begins with the lower plane and ends at the upper plane. The points in the grid represent the phonon number states. The transition probability is obtained by the collective projection measurements of two phonon modes (see SM.D.). (C) The fidelity analysis of the collective projection measurements. The fidelity of measuring the transition probability to the state $\mathrm{\ket{n_{X'},n_{Y'}}}$ is experimentally examined from $\mathrm{\ket{0,0}}$ to $\mathrm{\ket{9,9}}$. The fidelity is achieved by applying the measurement sequence twice, as starting from $\mathrm{\ket{0,0}}$ to $\mathrm{\ket{n_{X'},n_{Y'}}}$, and then bringing back to $\mathrm{\ket{0,0}}$. The square root of the remained population represents the fidelity.}
\label{Detection}
\end{figure}

Fig.~\ref{Detection}(A) depicts a scheme for reconstructing the spectrum at zero Kelvin from the output measurements of the trapped-ion simulator, the transition intensities from the ground state to the excited states are aligned according to the transition frequencies. Fig.~\ref{Detection}(B) illustrates the transition between the two-dimensional Fock spaces resulting from the two-dimensional harmonic oscillators. %\note{Joon: I think Fig3B is redundant but I leave it to you.} 
%In our simulation of SO$_2$, the Fock state is identified as $\mathrm{\ket{n_X,n_Y}}$, where $\mathrm{n_X}$, $\mathrm{n_Y}$ are the phonon numbers of motional $\mathrm{X}$ and $\mathrm{Y}$ modes. 
Finally, We perform the collective quantum-projection measurement of the final state $\mathrm{\ket{n_X,n_Y}}$ advanced from the measurement scheme of Ref. \cite{Um16,NOON2016}: first, we transfer the population of a target state $\mathrm{\ket{n_X,n_Y}}$ to the $\mathrm{\ket{0,0}}$ state by a sequence of $\pi$-pulse transitions. Then, we measure the state population by applying three sequential  fluorescence-detections combined with the uniform red sideband technique (see SM.D. for the detailed information). Our quantum projection measurement is limited by the imperfection of the state transfer and the fluorescence-detection efficiency (see SM.E.). We plot the fidelity of the collective projection measurement of $\mathrm{\ket{n_X,n_Y}}$ state in Fig.~\ref{Detection}(C). Based on the fidelity analysis, We perform measurement-error corrections for the experimental raw data (see SM.E. for the detailed information).

Furthermore, We simulate the photoelectron spectroscopies of SO$_2$ and SO$_2^-$, in which a single electron is  removed from the molecule during the photon absorption process, with our trapped-ion quantum simulator. Owing to the symmetry of the molecules, we use only two vibrational modes of sulfur dioxide, which show mixing of the vibration modes with respect to the final vibrational coordinates~\cite{Lee2009}, in our quantum simulation for the molecular spectra; the remaining vibrational mode does not contribute to the overall spectral shape because SO$_{2}$ does not deform along the remaining (non-totally-symmetric) vibrational mode during the vibronic transition.   
%~For the photoelectron spectroscopy of SO$_2$, $\boldsymbol{\delta} = (-0.026, 1.716)$ of displacement $\hat{D}$, $\boldsymbol{\zeta} = (0.288, -0.204)$ of squeezing operation $\hat{S}$, $\theta = 0.189$ of rotation $\hat{R}$, and $\boldsymbol{\zeta'} =( 0.317, -0.093)$ of inverse squeezing operation $\hat{S}^{\dagger}$ has been performed by following Eq.~\eqref{eq:Doktorov3}. We note that the non-subscript operators $\hat{D}$, $\hat{S}$ and $\hat{R}$ are specially used for the two-mode situation. 

\begin{figure}[htbp]
\centering
\includegraphics[width=\textwidth]{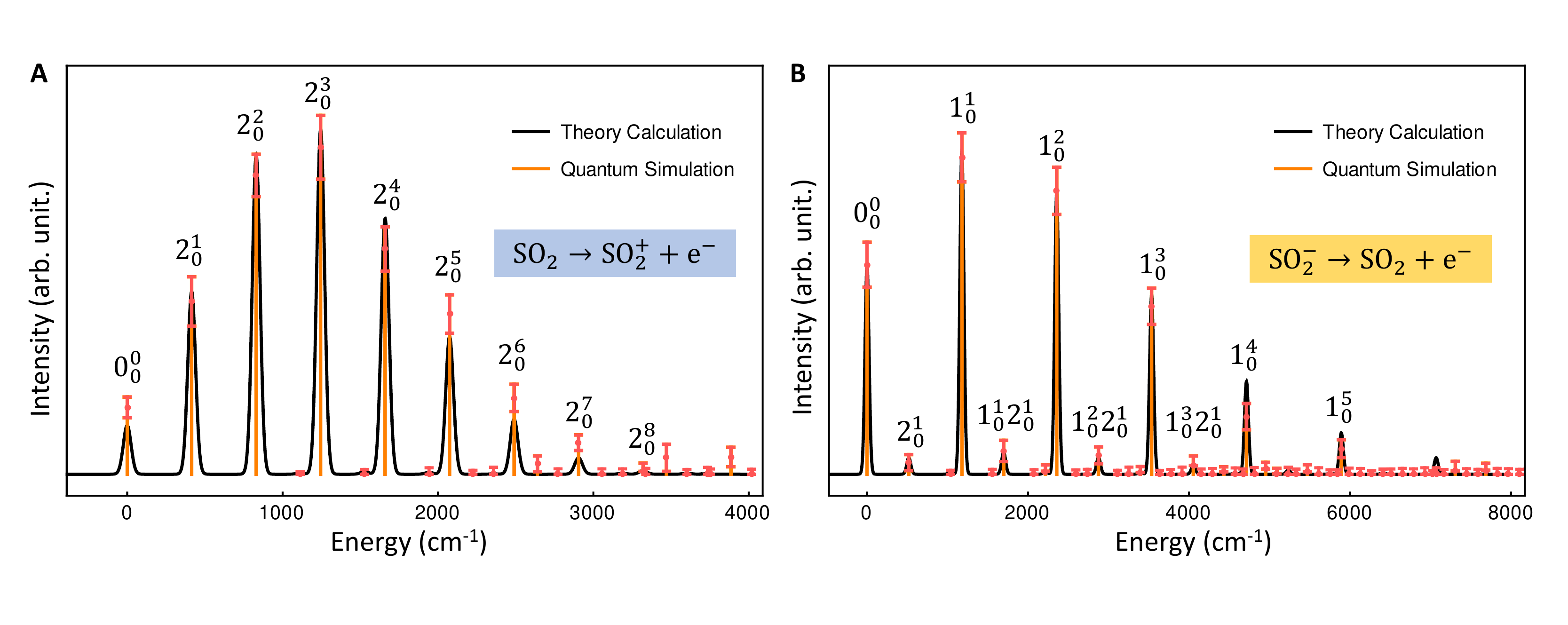}
\caption{\textbf{Trapped-ion simulation of photoelectron spectra of SO$_2$ and SO$_2^-$ with  measurement-error correction.}  
The two vibrational frequencies of harmonic potential for SO$_2^+$, SO$_2$ and SO$_2^-$ are (1112.7, 415.0), (1178.4, 518.9) and (989.5, 451.4) cm$^{-1}$, respectively. (A) The displacement vector $\boldsymbol{\delta}$ is (-0.026, 1.716); rotation angle $\theta$ is $0.189$ and (B) The displacement vector $\boldsymbol{\delta}$ is (1.360, -0.264); rotation angle $\theta$ is $0.065$. The theoretical lines are intentionally broadened by convoluting with a Gaussian function with the width of 50 cm$^{-1}$ \cite{Lee2009} for the comparison. Here $N^{i}_{0}$ denotes the $i$-phonon excitation on $N$-th mode from the vibrational ground state $\ket{0}$, and accordingly, $0^{0}_{0}$ located at the off-set energy $\omega_{0-0}=0$.
} 
\label{Spectroscopy}
\end{figure}

Fig.~\ref{Spectroscopy} presents the photoelectron spectra, SO$_{2}\rightarrow~$SO$_{2}^{+}$ and SO$_{2}^{-}\rightarrow~$SO$_{2}$ obtained from our trapped-ion quantum simulation, this is compared with the theoretical classical calculations.      
Fig.~\ref{Spectroscopy} shows good agreement between the theory calculations and the trapped-ion simulations of the two photoelectron processes of sulfur dioxide, the required molecular parameters are described in the figure caption. In Fig.~\ref{Spectroscopy}(A), the photoelectron spectrum of SO$_2$ is dominated by $\omega'_2$ transitions due to the significant large displacement along the second mode: $\boldsymbol{\delta} = (-0.026, 1.716)$. The photoelectron spectrum of SO$_2^-$ in Fig.~\ref{Spectroscopy}(B) shows tiny combination bands of the first and second modes regardless of the dominant contribution of the first mode ($\boldsymbol{\delta} =(1.360, -0.264)$). We note that the observation of the tiny band combinations indicates reliable performance of the trapped-ion simulation. %\note{Joon: I don't understand the last sentence after the semicolon.}

As the first demonstration of quantum simulation for the molecular vibronic spectroscopy, our trapped-ion device shows an excellent performance at a small-scale after the error-correction scheme in SM.E. In the near future, we expect the many modes implementation for a large-scale molecular simulation, where the multi-modes can be mapped to the local vibrational modes or collective normal modes of many ions in a single trap. The demonstrated quantum optical operations through a single ion will be directly applied for the large-scale simulation.   %\note{Joon: Please give one or two sentences in supporting the large-scale implementation in the near future.}
This would be useful in concluding the quantum supremacy of boson sampling with the Gaussian states. The molecular simulations in trapped-ion devices and the real molecular spectroscopic signals can be compared as a certification protocol for large-scale Gaussian boson sampling, which cannot be verified classically because of the \#P-hardness~\cite{rahimi2015,huh2015,huh2016}. 

%\bibliographystyle{Science}
%\bibliography{scibib}

\paragraph*{Acknowledgments:}
This work was supported by the National Key Research and Development Program of China under Grants No.
2016YFA0301900 (No. 2016YFA0301901), the National Natural Science Foundation of China Grant 11374178 and 11574002. JH acknowledges the supports by Basic Science Research Program through the National Research Foundation of Korea (NRF) funded by the Ministry of Education, Science and Technology (NRF-2015R1A6A3A04059773), the ICT R\&D program of MSIP/IITP [1711028311] and Mueunjae Institute for Chemistry (MIC) postdoctoral fellowship.

%\end{multicols}

\newpage

\newpage

\newpage

\newpage

\newpage
\section*{Supplementary Materials: Quantum simulation of molecular spectroscopy in trapped-ion device}

\author
{Yangchao Shen$^{1}$, Joonsuk Huh$^{2*}$, Yao Lu$^{1}$, Junhua Zhang$^{1}$, \\
Kuan Zhang$^{1}$, Shuaining Zhang$^{1}$ and Kihwan Kim$^{1\dagger}$\\
\\
\normalsize{$^{1}$Center for Quantum
Information, IIIS, Tsinghua University, Beijing, P. R. China,}\\
\normalsize{$^{2}$Department of Chemistry, Sungkyunkwan University, Suwon 440-746, Korea}\\
\\
\normalsize{To whom correspondence should be addressed;} 
\\
\normalsize{E-mails:  $^\ast$joonsukhuh@gmail.com and $^\dagger$kimkihwan@mail.tsinghua.edu.cn}
}

\maketitle

\setcounter{figure}{0}    
\renewcommand{\thefigure}{S\arabic{figure}}
\setcounter{equation}{0}
\renewcommand{\theequation}{S\arabic{equation}}

\subsection*{A. Quantum optical operators}
We define, herein, the quantum optical operators we have used in our work. 
$\boldsymbol{a}$ and $\boldsymbol{a^\dag}$ are $N$-dimensional column vectors for the bosonic annihilation and creation operators, respectively. That is,  
\begin{align}
\boldsymbol{a} \equiv (a_1, a_2, ... , a_N)^{\rm T} ~,~ \boldsymbol{a^\dag} \equiv (a_1^\dag, a_2^\dag, ... , a_N^\dag)^{\rm T}
\end{align}
where $[a_i,a_j^{\dagger}]=\delta_{ij}$.

The $N$-mode displacement operator is defined as below with the displacement vector $\boldsymbol{\delta}=(\delta_1, \delta_2, ..., \delta_N)$,
\begin{align}
\hat{D}_N(\boldsymbol{\delta})={\rm exp} \{  \boldsymbol{\delta a^\dag}-  \boldsymbol{\delta^* a }\}.
\end{align} 

The $N$-mode squeezing operator is defined as below with the squeezing parameter matrix $\boldsymbol{\zeta}={\rm diag} (\zeta_1, \zeta_2, ..., \zeta_N)$. 
\begin{align}
\hat{S}_N(\boldsymbol{\zeta})={\rm exp} \{\frac{\boldsymbol{a^{\mathrm{T}} \zeta^\dag a}}{2} - \frac{\boldsymbol{(a^\dag)^{\mathrm{T}}\zeta a^\dag}}{2}\}.
\end{align}

The $N$-mode rotation operator is defined as below with a unitary matrix $\boldsymbol{U}$, 
\begin{align} 
\hat{R}_N(\boldsymbol{U})={\rm exp} \{ \boldsymbol{ (a^\dag)^{\mathrm{T}}} \ln (\boldsymbol{U}) \boldsymbol{a} \}.
\end{align}

\subsection*{B.~Experimental parameters for quantum optical operations}
We present, here, the parameters used in the trapped-ion device for the quantum optical operations. 
The displacement operator with two modes is rewritten as follows, 
\begin{align}
\hat{D}_2(\boldsymbol{\delta}) = \hat{D} { \begin{pmatrix} {\delta_{\rm X}},{\delta_{\rm Y}} \end{pmatrix}} =   \mathrm{exp} \{  {\delta_{\rm X}} a^\dag_{\rm X}- {\delta^{*}_X} a_{\rm X} \} \mathrm{exp} \{ {\delta_{\rm Y}} a^\dag_{\rm Y}- {\delta^{*}_{\rm Y}} a_{\rm Y} \}. \label{Ddelta}
\end{align}
As seen in Eq. \ref{Ddelta}, the displacement operations of the X and Y modes can be implemented independently. 

The squeezing operator with the two mode parameter $\boldsymbol{\zeta}={\rm diag}(\ln \sqrt{\omega_1},\ln \sqrt{\omega_2})={\rm diag}(\zeta_{\rm X},\zeta_{\rm Y})$ can be rewritten as follows, 
\begin{align}
\hat{S}_2(\boldsymbol{\zeta}) = \hat{S}({\rm diag}(\zeta_{\rm X},\zeta_{\rm Y})) = \mathrm{exp} \{ \frac{\zeta_{\rm X}}{2} (a_{\rm X} a_{\rm X}-a_{\rm X}^\dag a_{\rm X}^\dag) \}  \mathrm{exp} \{ \frac{\zeta_{\rm Y}}{2} (a_{\rm Y} a_{\rm Y} - a_{\rm Y}^\dag a_{\rm Y}^\dag)\}.
  \label{squ2equ}
\end{align}
In the trapped-ion experiment, the squeezing operations are limited to the range of $\zeta_{\rm X}(\zeta_{\rm Y}) \leq 4$ in Eq.~\ref{squ2equ}. Since $\hat{U}_\mathrm{Dok}$ involves the squeezing and inverse squeezing operations, we can freely rescale the squeezing parameters with a single arbitrary constant: we rescale the squeezing parameters by a factor of 1/25, $(\zeta_{\rm X}, \zeta_{\rm Y}) = (\ln (\sqrt{\omega_1}/25),\ln (\sqrt{\omega_2}/25))$ and $(\zeta_{\rm X}', \zeta_{\rm Y}') = (\ln (\sqrt{\omega_1'}/25),$$\ln (\sqrt{\omega_2'}/25))$. The rescaling parameter is canceled after the two squeezing operations. 

The two mode rotation operation can be written simply with a rotation angle $\theta$, 
\begin{align}
\hat{R}_2(\boldsymbol{U})=\hat{R}(\theta)=e^{\theta(\hat{a}_{\rm X}^\dag \hat{a}_{\rm Y}-\hat{a}_{\rm X}\hat{a}_{\rm Y}^\dag)}
\end{align}
where $\boldsymbol{U}=\begin{pmatrix} \cos \theta & \sin \theta \\ -\sin \theta & \cos \theta \end{pmatrix}$ becomes the unitary rotation matrix. The rotation angle $\theta$ is controlled by Raman laser beams in the trapped-ion simulation.
  
\begin{table}[!htbp]
\centering
\caption{Parameters for the trapped-ion simulation of 
SO$_2$$\rightarrow$SO$_2^{+}$ and SO$_2^-$$\rightarrow$SO$_2$}.\label{SO2parameters}
\begin{tabular}{p{3cm}<{\centering}p{5cm}<{\centering}p{5cm}<{\centering}}
\toprule
  & SO$_2$$\rightarrow$SO$_2^{+}$& SO$_2^-$$\rightarrow$SO$_2$\\
\midrule
$\delta_{\rm X}, \delta_{\rm Y}$ &(-0.026, 1.716)& (1.360, -0.264)\\
$\omega_1', \omega_2'$& (1112.7, 415)& (1178.4, 518.9)\\
$\zeta_{\rm X}', \zeta_{\rm Y}'$& (0.288, -0.204)& (0.317, -0.093)\\
$\boldsymbol{U}$&$\begin{pmatrix} 0.982 & 0.188 \\ -0.188 & 0.982 \end{pmatrix}$&$ \begin{pmatrix} 0.998 & 0.065 \\ -0.065 & 0.998  \end{pmatrix}$\\
$\theta$&0.1892&0.065\\
$\omega_1, \omega_2$& (1178.4, 518.9)& (989.5, 451.4)\\
$\zeta_{\rm X}, \zeta_{\rm Y}$& (0.317, -0.093)& (0.229, -0.162)\\
\bottomrule
\end{tabular}
\end{table}

\subsection*{C.~Quantum optical operations in trapped-ion system}
\begin{figure}[htbp]
\centering
\includegraphics[width=0.6\textwidth]{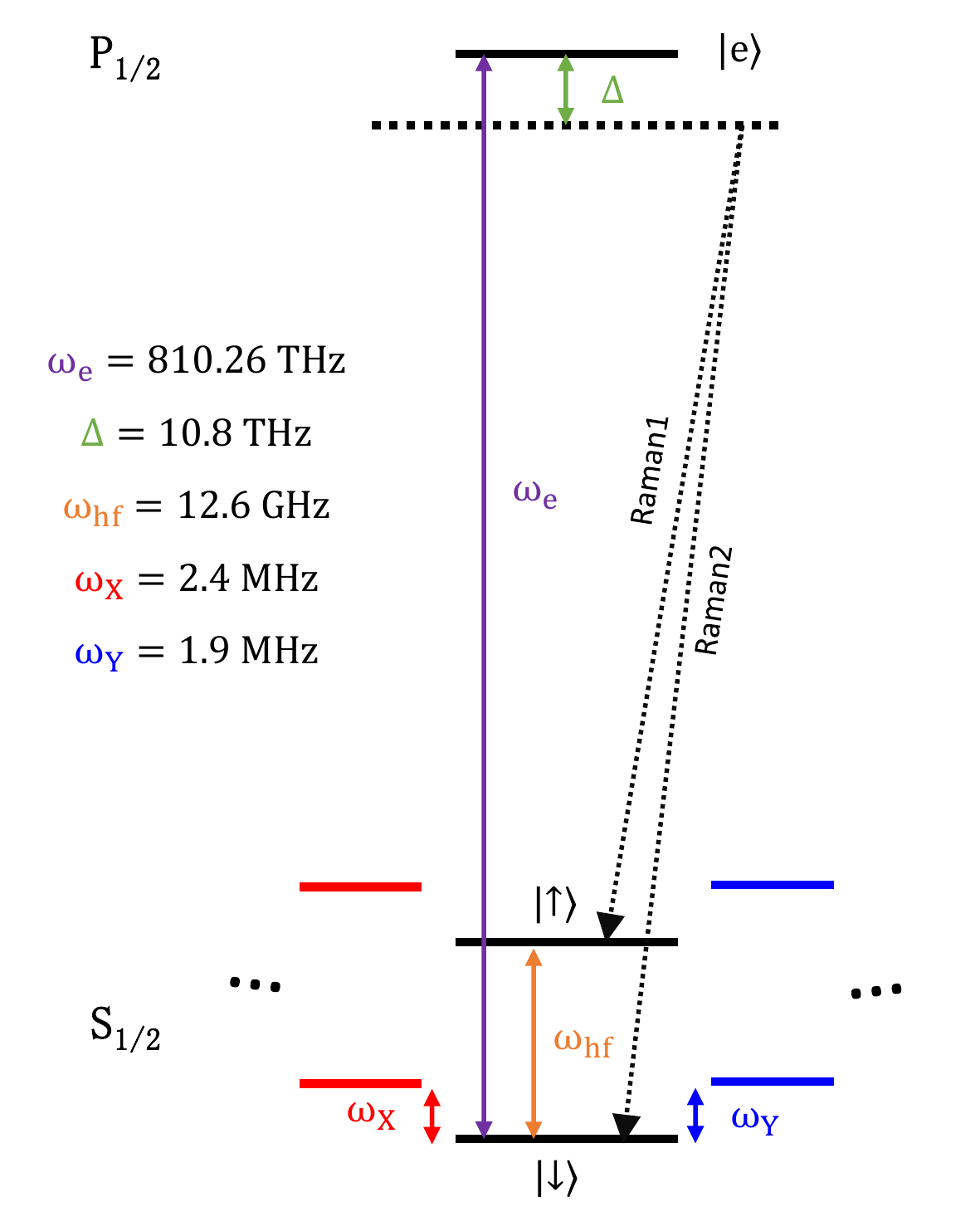}
\caption{\textbf{Energy level diagram of $^{171}\mathrm{Yb}^+$}. The electronic levels ($\ket{\uparrow}, \ket{\downarrow}$) with the difference $\omega_{\mathrm{hf}}$, and the phonon levels of modes X and Y with the frequencies of $\omega_{\rm X}$ and $\omega_{\rm Y}$ are involved in the Raman process. The Raman process is illustrated as a carrier transition between the two electronic states, $\ket{\uparrow}$ and $\ket{\downarrow}$.}
\label{FigureS1}
\end{figure}

%Earlier work\cite{TrappedionRMP2003} showed how to manipulate trapped ion phonon space through laser atom interaction, including carrier, blue and red sideband operations. 
We implement the quantum optical operations ($\hat{D}$, $\hat{S}$ and $\hat{R}$) via controlling Raman laser beams. %\note{Here, we briefly describe the controlling process please see Ref.[???] for the details.-> no reference that shows the exactly same discussion. } 
Figure~\ref{FigureS1} shows the energy diagram of a trapped $\mathrm{^{171}Yb^+}$. The two levels in hyperfine structure of $^{2}\mathrm{S}_{1/2}$ manifold is usually used to realize a qubit, which are denoted as $\ket{\downarrow} \equiv \ket{\mathrm{F=0,m_F=0}}$ and $\ket{\uparrow} \equiv \ket{\mathrm{F=1,m_F=0}}$. The red (mode X) and blue (mode Y) lines stand for the motional degrees of freedom. The Raman process is implemented via the virtual energy level, which is $\Delta$ detuned from $\mathrm{P_{1/2}}$ level, $\ket{e}$. Our quantum optical operations can be implemented by applying different frequencies of Raman laser beams.

To show how we can implement the quantum optical operations with the Raman laser beams, we start from a light-matter Hamiltonian. 
Based on the energy level diagram of $^{171}\mathrm{Yb}^+$ Figure~\ref{FigureS1}, the Hamiltonian for the whole system (atom and laser) can be written as,  
\begin{align}
H&=\dfrac{\hbar \omega_{\rm hf}}{2} \sigma_{\rm Z} +\hbar \omega_{\rm e} |e\rangle\langle e|+ \hbar \omega_{\rm X} (a_{\rm X}^\dagger a_{\rm X}+\dfrac{1}{2}) + \hbar \omega_{\rm Y} (a_{\rm Y}^\dagger a_{\rm Y}+\dfrac{1}{2})\nonumber\\
&+ \sum_{j=1}^2 \dfrac{\hbar g_j}{2}(e^{i (\vec{k}\cdot\vec{r}-\omega_j t+\phi_j)}+{\rm h.c.})(\ket{\uparrow}\bra{e}+\ket{\downarrow}\bra{e}+ \rm h.c.),
\end{align}
where $\vec{k_j}\cdot\vec{r} = k_{{\rm X}j} \hat{e}_{\rm X} + k_{{\rm Y}j} \hat{e}_{\rm Y}$ with the condition of $ k_{{\rm X}1} = -k_{{\rm X}2}$ and $ k_{{\rm Y}1} = -k_{{\rm Y}2}$ for the contour-propagating geometry of Raman laser beams. $g_j$ represents the strength of the dipole coupling  between $\ket{\uparrow},\ket{\downarrow}$ and $\ket{e}$. Here $j=1,2$ stand for the Raman beam.
%\note{Joon: Define g$_{j}$ either in here or in the figure}

\begin{figure}[htbp]
\centering
\includegraphics[width=\textwidth]{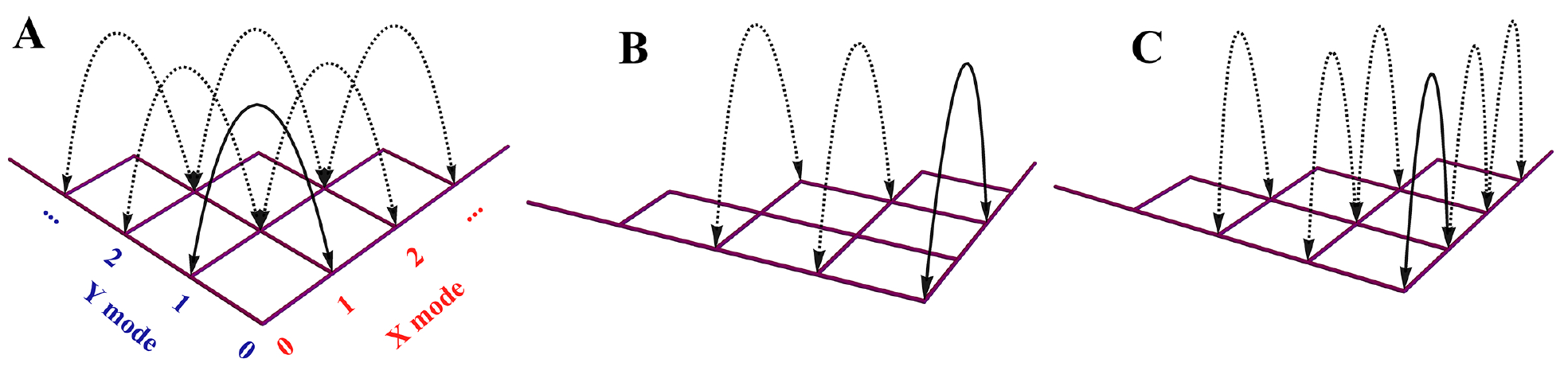}
\caption{{\bf The implementation and the effects of the quantum optical operations in the trapped-ion system.} The Hilbert space is composed of two phonon modes of X and Y. The quantum operations are implemented by setting different frequencies for the Raman laser beams. (A) Rotation operation $\hat{R}$ between X and Y modes. (B) Squeezing operation $\hat{S}$ and (C) coherent displacement operation $\hat{D}$ on X mode as an example. %\note{Joon: I don't understand this figure.}
}
\label{FigureRSD}
\end{figure}

After introducing the interaction picture to the Hamiltonian with respect to the system part, $H_0=\dfrac{\hbar \omega_{\rm hf}}{2} \sigma_{\rm Z} +\hbar \omega_{\rm e} |e\rangle\langle e|+ \hbar \omega_{\rm X} (a_{\rm X}^\dagger a_{\rm X}+\dfrac{1}{2}) + \hbar \omega_{\rm Y} (a_{\rm Y}^\dagger a_{\rm Y}+\dfrac{1}{2})$; and by setting two Raman laser beam frequencies as $\omega_1 = \omega_{\rm e} - \Delta,$ $\omega_2 = \omega_{\rm e} - \Delta + (\omega_{\rm X} -\omega_{\rm Y})$, which is represented by the configuration shown in Fig.~\ref{FigureRSD}(A) ; finally, by using rotating wave and Lamb-Dicke approximation together with the experimental conditions of $\Delta\gg\omega_{\rm hf}\gg\omega_{\rm X},\omega_{\rm Y}\gg g_1,g_2$, we arrive at the Hamiltonian for the rotation operation $\hat{R}$ as, 
\begin{align}
H_{\hat{R}} = \dfrac{\hbar g_1 g_2\eta_{\rm X}\eta_{\rm Y}}{2\Delta}(a_{\rm X}^\dagger a_{\rm Y} e^{i\Delta\phi}+a_{\rm X} a_{\rm Y}^\dagger e^{-i\Delta\phi} ),
\end{align}
which leads to the rotation $\hat{R}$
\begin{align}
\hat{R}(\theta) = \exp\left\{-it\dfrac{\hbar g_1 g_2\eta_{\rm X}\eta_{\rm Y}}{2\Delta}(a_{\rm X}^\dagger a_{\rm Y} e^{i\Delta\phi}+a_{\rm X} a_{\rm Y}^\dagger e^{-i\Delta\phi} )\right\}.
\end{align}
Here $\theta=t\dfrac{\hbar g_1 g_2\eta_{\rm X}\eta_{\rm Y}}{2\Delta}$, $\Delta\phi=\phi_2-\phi_1$ is the relative phase between two Raman beams, and $\eta_\mathrm{X}=0.117$, $\eta_\mathrm{Y}=0.132$ are the Lamb-Dicke parameter for the Raman process. In the experiment, we set $\dfrac{\hbar g_1 g_2\eta_{\rm X}\eta_{\rm Y}}{2\Delta}= 0.006$ rad $\mu s^{-1}$. 

Similarly, we can perform the squeezing operation $\hat{S}$ of single motional mode (here, mode X as an example) by setting two Raman laser frequencies as $\mathrm{\omega_1 = \omega_e - \Delta,~~ \omega_2 = \omega_e - \Delta + 2 \omega_X }$, which leads the configuration shown in Fig.~\ref{FigureRSD}(B), 
\begin{align}
\hat{S}(\zeta_{\rm X},0) =\exp \left\{-it \dfrac{\hbar\Omega_1\Omega_2\eta_{\rm X}^2}{2\Delta}(a_{\rm X}^\dagger a_{\rm X}^\dagger e^{i\Delta\phi}+a_{\rm X} a_{\rm X} e^{-i\Delta\phi} )\right\},
\end{align}
where $\zeta_{\rm X}=t\dfrac{\hbar\Omega_1\Omega_2\eta_{\rm X}^2}{\Delta}$. In the experiment, we set $\dfrac{\hbar\Omega_1\Omega_2\eta_{\rm X}^2}{\Delta}=0.006~\mu s^{-1}$. 

The displacement operation $\hat{D}$ of single mode (here, mode X as an example) by setting two Raman laser frequencies $\mathrm{\omega_1 = \omega_e - \Delta,}$ $\mathrm{\omega_2 = \omega_e - \Delta + \omega_X }$, which leads the configuration shown in Fig.~\ref{FigureRSD}(C),  
\begin{align}
\hat{D}(\delta_{\rm X},0) =\exp\left\{ -it\dfrac{ \hbar\Omega_1\Omega_2\eta_{\rm X}}{2\Delta}(a_{\rm X}^\dagger e^{i\Delta\phi}+a_{\rm X} e^{-i\Delta\phi})\right\},
\end{align}
where $\delta_{\rm X}=t\dfrac{\hbar\Omega_1\Omega_2\eta_{\rm X}}{2\Delta}$. In the experiment, we set $\dfrac{\hbar\Omega_1\Omega_2\eta_{\rm X}}{\Delta}=0.066~\mu s^{-1}$. 

%Figure~\ref{FigureRSD} illustrates the implementations of the quantum optical operations pictorially and the effects of applying them in the Hilbert space formed by X and Y phonon modes. \note{Joon: I don't understand this sentence and the figure. Please explain more or improve the description.} 

%\note{Shen, Please proved a table like Table 1 for the parameters in the Hamiltonians in S9, S10, S11 -> now we do not think we need the table since we put the experimental values of them} 

\subsection*{D.~Method for collective projection measurements}

\begin{figure}[htbp]
\centering
\includegraphics[width=\textwidth]{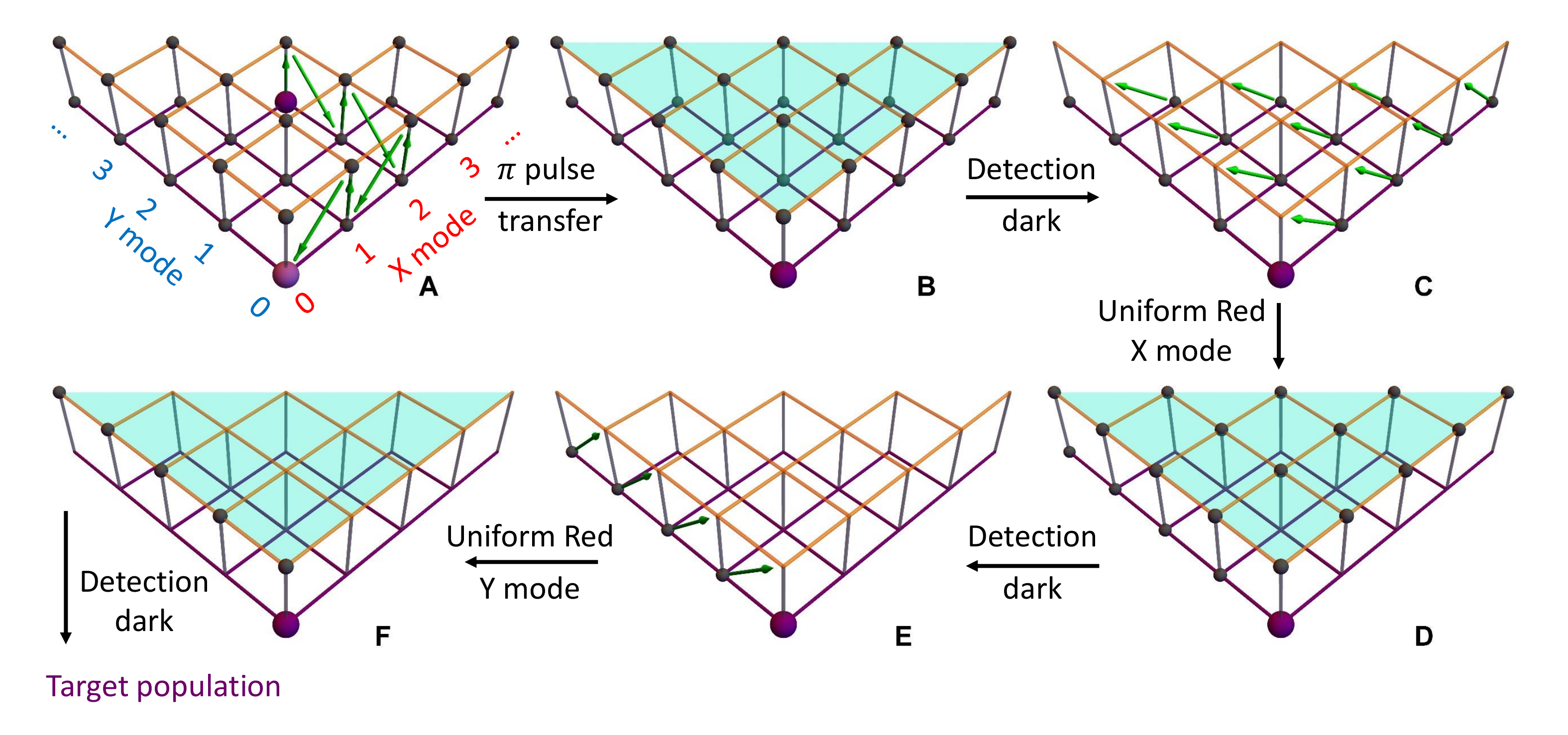}
\caption{{\bf Detection method for the state $\ket{\downarrow, n_{\rm X}=2, n_{\rm Y}=2}$}. The lower and the upper layers represent the internal states of $\ket{\downarrow}$ and $\ket{\uparrow}$. The internal states have no fluorescence and fluorescence, respectively, during the internal state detection.}  
\label{Detection_method}
\end{figure}
We explain in this section the pulse sequence for the detection of population in an arbitrary phonon state $\mathrm{\ket{\Sigma,n_X,n_Y}}$, where we indicate the internal qubit state $\Sigma$ ($\downarrow$ or $\uparrow$) of the phonon state ($\mathrm{\ket{n_X,n_Y}}$). 

The first step is to transfer the population in the target state $\mathrm{\ket{\downarrow,n_X,n_Y}}$ to $\mathrm{\ket{\downarrow,0,0}}$: it is performed by applying a sequence of $\pi$-pulse transitions, as shown in Fig.~\ref{Detection_method}(A), i.e., with the following steps, 
\begin{eqnarray*}
\mathrm{(A)}:~\mathrm{\ket{\downarrow,n_X,n_Y}} & \xrightarrow{\mathrm{\pi-Carrier}} & \ket{\uparrow,n_X,n_Y} \xrightarrow{\mathrm{\pi-Blue X}} \ket{\downarrow,n_X-1,n_Y}... \longrightarrow ... \ket{\downarrow,0,n_Y}\\ & \xrightarrow{\mathrm{\pi-Carrier}} & \ket{\uparrow,0,n_Y} \xrightarrow{\mathrm{\pi-Blue Y}} \ket{\downarrow,0,n_Y-1} ... \longrightarrow ... \mathrm{\ket{\downarrow,0,0}}
\end{eqnarray*}

The second step is to obtain the population in $\mathrm{\ket{\downarrow,0,0}}$ by using the sequence as shown in Fig.~\ref{Detection_method}(B-F).

(B): Apply the fluorescence detection and record the event $M_1$ of detecting photons or no photons.

(C): Apply a uniform red-transition on mode X, which transfers all the states of $\mathrm{\ket{\downarrow,n_{\rm X}>0,n_{\rm Y}}}$ to $\ket{\mathrm{\uparrow}}$ state.

(D): Apply the fluorescence detection and record the event $M_2$ of detecting photons or no photons.

(E): Apply a uniform red-transition on Y mode, which transfers all the states of $\mathrm{\ket{\downarrow,n_{\rm X},n_{\rm Y}>0}}$ to $\ket{\mathrm{\uparrow}}$ state.

(F): Apply the fluorescence detection and record the event $M_3$ of detecting photons or no photons.

In the above multiple-detection stages, there are four situations for the recorded data $M_1M_2M_3$ $$\mathrm{\{B \forall \forall, DB \forall,DDB,DDD\}} \rightarrow \{P_1,P_2,P_3,P_4\}.$$ 
Here, $\mathrm{D}$ means detecting no photons, $\mathrm{B}$ means detecting photons, $\mathrm{\forall}$ stands for both situations. Typically, we repeat the experiments for 2000 times to get the probability for each case noted as ${P_1,P_2,P_3,P_4}$. The population of the target state is the probability of case $P_4$.

Within the above collective projection measurements, Fig.~3(C) shows the experimentally measured result for the fidelity of the detection sequence of an arbitrary state $\mathrm{\ket{n_X,n_Y}}$, noted as $F_{D.M}$. The infidelity mainly comes from the imperfection of $\pi$-pulse and uniform red-transition on X and Y mode. 

\subsection*{E.~Measurement-error corrections for the experimental raw data }
We mainly consider two error sources to correct the experimental raw data: 
i) the inefficiency of fluorescence detection of internal states;
ii) the infidelity of the collective projection measurement discussed in section D.

Our fluorescence detection can distinguish the internal states $\ket{\mathrm{\uparrow}}$ and $\ket{\mathrm{\downarrow}}$ with the corresponding detection fidelities are $ \eta_{\mathrm{\uparrow \rightarrow \uparrow}}~(97.2\%) $ state and $\eta_{\mathrm{\downarrow \rightarrow \downarrow}}~(99.3\%)$ for state, respectively. To correct this inefficiency, we use the value of $P_4$, which is obtained by 1-($P_1 + P_2 + P_3$). The real population ($P_R$) of detecting photons scattered from the $\ket{\mathrm{\uparrow}}$ state is not exactly same to the measured population ($P_M$). The relation between them is given as $P_M=P_R ~\eta_{\uparrow \rightarrow \uparrow}+(1-P_R) (1-\eta_{\downarrow \rightarrow \downarrow})$, thus
$$P_R\equiv \mathrm{Corr}(P_M)=\frac{P_M-(1-\eta_{\downarrow \rightarrow \downarrow}) }{\eta_{\downarrow \rightarrow \downarrow}+\eta_{\uparrow \rightarrow \uparrow}-1}$$
For the correction of the second part, as discussed in the section C, we have to include the fidelity $F_{D,M}$.

In order to correct the raw experimental data, we consider these two imperfections. For the experiment raw data, our corrected data is written accordingly as,  
\begin{align}
P'_4=\frac{1-\mathrm{Corr}(P_1+P_2+P_3)}{F_{D.M}}
\end{align}

\begin{figure}[htbp]
\centering
\includegraphics[width=\textwidth]{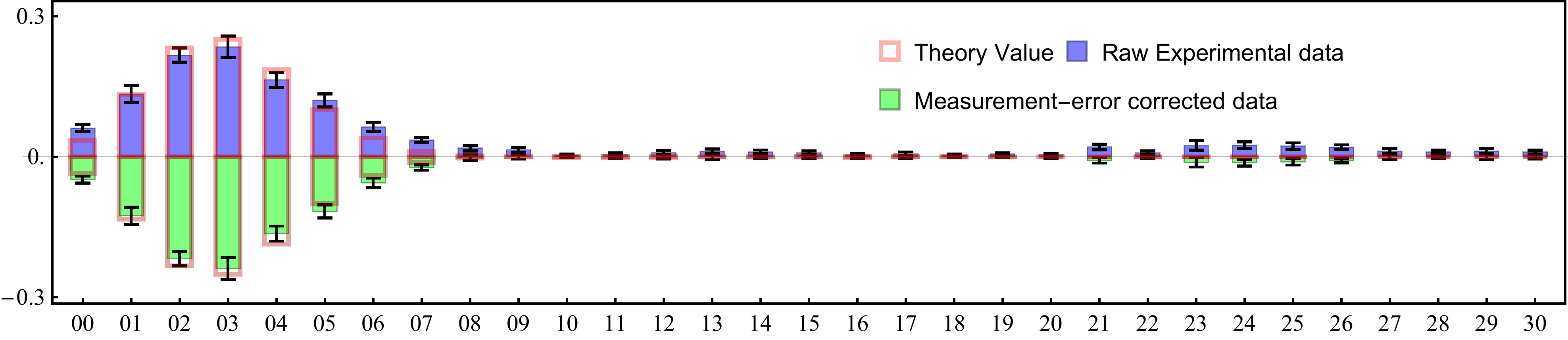}
\caption{{\bf Comparison between the raw and the corrected experimental data for the spectroscopy of SO$_2$$\rightarrow$SO$_2^{+}$.} The horizontal axis is the Fock state  $\mathrm{\ket{n_{X'},n_{Y'}}}$ and  the vertical axis is the transition intensity to the state from the $\ket{0,0}$ state.}
\label{FigureS4}
\end{figure}

Figure \ref{FigureS4} compares the raw experimental data and corrected data for the photoelectron spectroscopy of SO$_2$.

\end{document}